\documentclass[submission,copyright,creativecommons]{eptcs}
\providecommand{\event}{VORTEX 2018} 
\usepackage{underscore}           
\usepackage{textcomp}
\usepackage{stmaryrd}
\usepackage{amsthm}
\usepackage{amsmath}
\usepackage{amssymb}
\usepackage[dvipsnames]{xcolor}

\usepackage{booktabs}   
\usepackage{subcaption} 

\usepackage[draft]{fixme}
\FXRegisterAuthor{vs}{evs}{VS$\!\!$} 
\FXRegisterAuthor{vp}{VP}{\color{violet}VP$\!\!$}
\FXRegisterAuthor{sj}{SJ}{\color{orange}SJ$\!\!$}

\fxsetup{targetlayout=color}
\fxsetface{margin}{\tiny}
\fxsetup{envface=\tiny\it}  
\fxsetup{inlineface=\tiny\bfseries}

\usepackage{todonotes}
\usepackage{listings}

\lstdefinelanguage{hfc}{
	basicstyle=\lst@ifdisplaystyle\small\fi\ttfamily,
        columns=flexible,
	frame=single,
        xleftmargin=3.5pt,
        xrightmargin=3.5pt,
	basewidth=0.5em,
	sensitive=true,
	morestring=[b]",
	morecomment=[l]{//},
	morecomment=[n]{/*}{*/},
	commentstyle=\color{OliveGreen},
	keywordstyle=\color{blue}\textbf, keywords={def}, otherkeywords={=>,||,\&\&},
	keywordstyle=[2]\color{red}\textbf, keywords=[2]{rep,nbr,nbrLocal,nbrRemote,if,let,in},
	keywordstyle=[3]\color{violet}, keywords=[3]{mux,dist,lag,sumHood,countHood,allHood,anyHood,minHood,maxHood,foldHood,nbrVector,min,myID,1st,2nd},
	keywordstyle=[4]\color{orange}\textbf, keywords=[4]{spawn,share},
	keywordstyle=[5]\color{blue}, keywords=[5]{false,true,infinity,null}
}
\newcommand{\fdtodoin}[1]{\todo[color=magenta!40, author=FD,inline]{#1}}
\newcommand{\vstodoin}[1]{\todo[color=magenta!40, author=VS,inline]{#1}}

\newcommand{\ID}{\iota}
\newcommand{\BNFcce}{{\bf ::=}}
\newcommand{\BNFmid}{\;\bigr\vert\;}

\newcommand{\PROGRAM}{\mathtt{P}}
\newcommand{\FUNCTION}{\mathtt{F}}
\newcommand{\e}{\mathtt{e}}
\newcommand{\fname}{\mathtt{d}}
\newcommand{\bname}{\mathtt{b}}
\newcommand{\xname}{\mathtt{x}}
\newcommand{\anyvalue}{\mathtt{v}}
\newcommand{\lvalue}{\ell}
\newcommand{\fvalue}{\phi}
\newcommand{\funvalue}{\mathtt{f}}

\newcommand{\dc}{\mathtt{c}}
\newcommand{\dcOf}[2]{#1(#2)}
\newcommand{\defK}{\mathtt{def}}
\newcommand{\nbrK}{\mathtt{nbr}}
\newcommand{\repK}{\mathtt{rep}}
\newcommand{\ifK}{\mathtt{if}}
\newcommand{\toSymK}{\mathrm{\texttt{=>}}}
\newcommand{\envmap}[2]{#1\mapsto #2}
\newcommand{\neigh}{\rightsquigarrow}
\newcommand{\denotexp}[3]{\mathcal{E}\llbracket {#1} \rrbracket_{#2}^{#3}}
\newcommand{\bsopsem}[5]{#1;#2;#3\vdash #4\Downarrow #5}
\newcommand{\nettran}[3]{#1\xrightarrow{#2} #3}
\newcommand{\act}{\textit{act}}
\newcommand{\envact}{\textit{env}}

\newcommand{\locationOf}{\textit{loc}}

\theoremstyle{definition}
\newtheorem{definition}{Definition}
\theoremstyle{remark}
\newtheorem{remark}{Remark}
\newtheorem{example}[remark]{Example}

\title{On Distributed Runtime Verification \\ by Aggregate Computing}
\author{Giorgio Audrito
 \qquad\qquad Ferruccio Damiani
\institute{University of Turin, Italy}
\email{\quad giorgio.audrito@gmail.com \quad\qquad ferruccio.damiani@unito.it}
\and
Volker Stolz
\institute{Western Norway University of Applied Sciences, Norway}
\email{Volker.Stolz@hvl.no}
\and
Mirko Viroli
\institute{University of Bologna, Italy}
\email{mirko.viroli@unibo.it}
}
\def\titlerunning{On Distributed Runtime Verification by Aggregate Computing}
\def\authorrunning{G. Audrito, F. Damiani, V.  Stolz \& M. Viroli}

\begin{document}
\maketitle

\begin{abstract}
Runtime verification is a computing analysis paradigm based on observing a system at runtime (to check its expected behaviour) by
means of monitors generated from formal specifications. Distributed runtime verification is runtime verification
 in connection with distributed systems: it comprises both monitoring of distributed systems and using distributed systems for monitoring.
Aggregate computing is a programming paradigm based on a reference computing machine that is the aggregate collection of devices
 that cooperatively carry out a computational process: the details of behaviour, position and number of devices are largely abstracted away,
 to be replaced with a space-filling computational environment. 
In this position paper we argue, by means of simple examples, that aggregate computing is particularly well suited for implementing  distributed monitors.
Our aim is to foster further research on  how to  generate aggregate computing monitors from suitable formal specifications.
\end{abstract}

\section{Introduction}\label{sec:Introduction}

Runtime verification is a computing analysis paradigm based on observing a system at runtime (to check its expected behaviour) by
means of monitors generated from formal specifications. Distributed runtime verification is runtime verification
 in connection with distributed systems: it comprises both monitoring of distributed systems and using distributed systems for monitoring.
Being a verification technique, additionally, runtime verification promotes the generation of monitors from formal specifications, so as to precisely state the properties to check as well as providing formal guarantees about the results of monitoring.
Distribution is hence a particularly challenging context in verification, for it requires to correctly deal with aspects such as synchronisation, faults in communications, possible lack of unique global time, and so on.
Additionally, the distributed system whose behaviour is to be verified at runtime could emerge from modern application scenarios like the Internet-of-Things (IoT), Cyber-Physical Systems (CPS), or large-scale Wireless Sensor Networks (WSN). In this case additional features are to be considered, like openness (the set of nodes is dynamic), large-scale (a monitoring strategy may need to scale from few units up to thousands of devices), and interaction locality (nodes may be able to communicate only with a small neighbourhood, though the property to verify is global).
So, in the most general case, distributed runtime verification challenges the way in which one can express properties on such dynamic distributed systems, can express flexible computational tasks, and can reason about compliance of properties and corresponding monitoring behaviour.

In this paper, we argue that a promising approach to address these challenges can be rooted on the computational paradigm of aggregate computing \cite{BPV-COMPUTER2015}, along with the field calculus language \cite{DVB-SCP2016}.
Aggregate computing promotes a view of distributed systems as a conceptually single computing device, spread throughout the (physical or virtual) space in which nodes are deployed.
At the paradigm level, hence, this view promotes the specification (construction, reasoning, programming) of global-level computational behaviour, where the interaction of individuals are essentially abstracted away.
At the modelling level, the field calculus can be leveraged, which expresses computations as transformations of \emph{computational fields} (or \emph{fields} of short), namely, space-time distributed data structures mapping computational events (occurring at a given position of space and time) to computational values.
As an example, a set of temperature sensors spread over a building forms a field of temperature values (a field of reals), and a monitor alerting areas where the temperature was above a threshold for the last 10 minutes is a function from the temperature field to a field of Booleans.
Field calculus has a working implementation called \textsc{ScaFi} in the Scala programming language \cite{CV-PLMDC2016}, where field computations can be expressed by Scala functions (relying on a suitable API) and actors are generated to realise the distributed system.

The remainder of this paper is organized as follows: 
Section~\ref{sec:Background} 
provides the necessary  background; 
Section~\ref{sec:FieldCalculus} presents the field calculus; 
Section~\ref{sec:GeneralMonitor} discusses monitoring general distributed programs through field calculus;
Section~\ref{sec:FCMonitor} illustrates how field calculus programs can be instrumented with monitors;
and Section~\ref{sec:Conclusion}  concludes.

\section{Background}\label{sec:Background}

In this section we provide the necessary background on distributed runtime verification and  aggregate computing by  briefly discussing the literature  
outlining their role.

\subsection{Motivating examples} \label{ssec:examples}

We motivate the usage of aggregate computing together with runtime verification techniques through two examples that have been thoroughly studied in earlier literature: a crowd-evacuation scenario and a general communication channel.

In the first scenario, a program (not necessarily written through aggregate computing techniques) is used to manage evacuation of agents from a given area in case of an emergency. Ideally, in such critical situations correctness guarantees for a particular solution and its implementation would be needed. Since the guarantees that can be proved are usually not fully satisfactory, they can be fruitfully complemented with runtime monitors. As an example, we focus here on a simple ``per-agent'' property that we could monitor: \emph{two neighbour agents} (closer than 5m) \emph{should not have ``evacuation vectors'' that lead to a direct collision with each other} (for both agents, the evacuation vector is within 60\textdegree{} from the direction of the other agent). We can then instantiate an aggregate computing monitor on each agent, observing the local and neighbours' evacuation vectors, and flagging violations of this property as they occur.

The second scenario describes an aggregate computing solution to the well-known problem of establishing a shortest communication path between two nodes, while ensuring reliability through an imposed \emph{width} (cross-section size of the channel), which provides the desired redundancy and alternative routes. In this situation, we define a more interesting property that is not per-agent, but rather \emph{per-network}, and also shows how the monitor can feedback into the original program: for each cross-section, we require that it has at least \emph{min} width of alternative connections. If not, we demand the channel program to increase the width. Conversely, if all slices contain more than \emph{max} nodes, we shrink the channel to save computational power.


\subsection{Distributed runtime verification}\label{sec:DistributedRuntimeVerification}
Runtime verification is a lightweight verification technique concerned with observing the execution of a system with respect to a specification \cite{DBLP:journals/jlp/LeuckerS09}.
Specifications are generally trace- or stream-based, with events that are mapped to atomic propositions in the underlying logic of the specification language.
Popular specification languages include variations on the Linear Temporal Logic (LTL), and regular expressions, which can be effectively checked through finite automata constructions.
Events may be generated through state changes or execution flow, such as method calls.

In \emph{distributed} runtime verification, we lift this concept to distributed systems, where we find applications in the following areas \cite{Francalanza2018}:
\emph{(i)} observing distributed computations and expressiveness (specifications over the distributed systems), \emph{(ii)} analysis decomposition (coupled composition of system- and monitoring components), \emph{(iii)} exploiting parallelism (in the evaluation of monitors), \emph{(iv)} fault tolerance and \emph{(v)} efficiency gains (by optimising communication).
In Sections~\ref{sec:GeneralMonitor} and \ref{sec:FCMonitor}, we show how runtime verification can be applied to, or contribute to some of those areas.

Naturally, such lifting also affects the specification language.
Bauer and Falcone \cite{DBLP:journals/fmsd/0002F16} show a decentralised monitoring approach where disjoint atomic propositions in a global LTL property are monitored without a central observer in their respective components.
Communication overhead is shown to be lower than the number of messages that would need to be sent to a central observer.

Sen \emph{et al.} introduce PT-DTL \cite{SenPTDLTL2004} to specify distributed properties in a past time temporal logic. 
Sub-formulas in a specification are explicitly annotated with the node (or process) where the sub-formula should be evaluated.
Communication of results of sub-computation is handled by message passing.

Both approaches assume a total communication topology, i.e., each node can send messages to everyone in the system,
although causally unrelated messages may arrive in arbitrary order.

Going beyond linear-time properties, hyperproperties over a set of traces allow a richer expressivity \cite{DBLP:conf/rv/FinkbeinerHST17}.
In our setting, as each node is running the same program, we can understand such a set as consisting of traces from the individual nodes.
Further issues on (efficient) monitorabilty have been addressed by Aceto \emph{et al.} in \cite{Aceto2018Monitorability}.

\subsection{Aggregate computing}\label{sec:AggregateComputing}

The problem of finding suitable programming models for ensemble of devices has been the subject of intensive research---see e.g. the surveys \cite{SpatialIGI2013,Viroli-Et-al:COORDINATION-2018}: works as TOTA~\cite{tota} and Hood~\cite{hood} provide abstractions over the single device to facilitate construction of macro-level systems; GPL~\cite{coorephd} and others are used to express spatial and geometric patterns; Regiment~\cite{regiment} and TinyLime~\cite{Curino05mobiledata} are information systems used to stream and summarise information over space-time regions; while MGS~\cite{GiavittoMGS05} and the 
fixpoint approach in~\cite{DBLP:journals/corr/Lluch-LafuenteL16} provide general purpose space-time computing models.
Aggregate computing and the field calculus have then be developed as a generalisation of the above approaches, with the goal of defining a programming model with sufficient expressiveness to describe complex distributed processes by a functional-oriented compositional model, whose semantics is defined in terms of gossip-like computational processes.

Hence, \emph{aggregate computing} \cite{BPV-COMPUTER2015} aims at supporting reusability and composability of collective adaptive behaviour as inherent properties.
Following the inspiration of ``fields'' of physics (e.g., gravitational fields), this is achieved by the notion of \emph{computational field}  (simply called \emph{field}) \cite{tota}, defined as a global data structure mapping devices of the distributed system to computational values. 
Computing with fields means deriving in a computable way an output field from a set of input fields. This can be done at a low-level, by defining programming language constructs or general-purpose building blocks of reusable behaviour, or at a high-level by designing collective adaptive services or whole distributed applications---which ultimately work by getting input fields from sensors and process them to produce output fields to actuators. 

The \emph{field calculus} \cite{DVB-SCP2016,viroli:selfstabilisation} is a minimal functional language that identifies basic constructs to manipulate fields, and whose operational semantics can act as blueprint for developing toolchains to design and deploy systems of possibly myriad devices 
interacting via proximity-based broadcasts.
%
%
Recent works have also adopted this field calculus as a \emph{lingua franca} to investigate formal properties of resiliency to environment changes 
\cite{DiGamma,viroli:selfstabilisation}, and to device distribution \cite{BVPD-TAAS2017}.

\subsection{Deployment}\label{sec:Deployment}

A number of techniques exist to deploy runtime verification as part of or in parallel to an application to be subjected to runtime verification.
A high-level technique to monitor a JVM-based application is the use of aspect-oriented programming \cite{StolzB06}, which allows for an easy integration in terms of events:
this method allows to easily intercept actions of the main application and use them as input events for the step-wise evaluation of properties.
In addition, this approach can be used to inspect or sample the current state of the systems.
This does not necessarily have to mean that the runtime verification algorithm is executed in the context of an application, but this event-generation can also be used to generate stimuli to external runtime verification engines that are implemented for example with the help of rewriting logic.

In the setting of field calculus programs, such an integration is more straight-forward:
here, we do not need to establish a coupling between a target application and a runtime verification framework,
but rather have FC programs that implement runtime verification monitors along side applications written in that formalism.
As such, they use the same communication constructs to aggregate information from neighbours and trigger local actions.

As in the more traditional RV approaches for main-stream languages and systems, also here one can separate the implementation language from the specification language.
We take a first step and show how common safety properties can be expressed as field calculus programs.
Ideally, one would next strive for a specification language that resembles more a temporal logic with future or past operators, which is then translated into a field calculus program to monitor the property.

Taken as an approach to \textit{distributed} runtime verification, we note that the field calculus also brings infrastructure that tackle a challenge in truly distributed systems:
the dynamic nature of these systems with their varying number of participants and communication topology poses the challenge of reliability.
So far, the RV community has mostly considered systems with a fixed number of agents and a fixed topology where communication is either point-to-point, allowing for interesting schemes to convey partial information, or broadcast, where message loss is not taken into account.
See Basin \emph{et al.}'s work \cite{BasinKZ15} for a rare take on distributed runtime verification in the presence of communication delays and errors.

In field calculus, on the one hand one faces the same challenges, e.g. of establishing a global property across all agents.
On the other hand, the constructs and mechanism of the field calculus, do provide a solution in themselves and do not require another level of middleware:
a developer that is already familiar with the field calculus will naturally encode e.g.\  properties of resilience and awareness of network partitions into their specifications.

\section{The field calculus}\label{sec:FieldCalculus}

The  \emph{field calculus} \cite{DVB-SCP2016} is a minimal language to express aggregate computations over distributed networks of (mobile) devices, each asynchronously capable of performing simple local computations and interacting with a neighbourhood by local exchanges of messages. Field calculus provides the necessary mechanism to express and compose such distributed computations, by a level of abstraction that intentionally neglects explicitly management of synchronisation, message exchanges between devices, position and quantity of devices, and so on; while retaining Turing-universality for distributed computations \cite{a:fcuniversality}.

\subsection{The model of computation} \label{ssec:model}

In field calculus, a program $\PROGRAM$ is periodically and asynchronously executed on every device, according to the following cyclic schedule. The involved device $\ID$, every period $T_\ID$:
\begin{enumerate}
	\item
	perceives contextual information, which is formed by: data provided by sensors, local information stored in the previous round, and messages collected from neighbours while sleeping,\footnote{Older messages may be retained until a certain timeout expires, or newer messages are received.} the latter in the form of a \emph{neighbouring value} $\fvalue$---essentially a map from neighbours to values $\anyvalue$;
	\item
	evaluates the program $\PROGRAM$, considering as input the contextual information gathered as described above;
	\item
	the result of this computation is a data structure that is stored locally, broadcast to neighbours, and possibly fed to actuators;
	\item
	sleeps until it is awaken at the next activation.
\end{enumerate}
By repetitive execution of such computation rounds, across space (where devices are located) and time (when devices fire), a global behaviour emerges~\cite{viroli:selfstabilisation}, which can be fruitfully considered as occurring on the overall network of interconnected devices, modelled as a single aggregate machine equipped with a neighbouring-based topology relation. This process can be mathematically modelled through the notion of \emph{event}, which correspond to the instants when devices are activated and start this sequence (see \cite{a:fcuniversality,a:rtssgradient} for further details on events and their role in modelling distributed computations).

\begin{definition}[event \cite{a:rtssgradient}] \label{def:event}
  An \emph{event} $e$ is modelled by the pair $e=(\ID,t)$ such that $\ID$ is the identifier of the device where the event takes place, and $t$ is the time when the device $\ID$ is activated. The time stamp $t$ refers to the local clock of $\ID$.
\end{definition}

Events are partially ordered by the following relationship.

\begin{definition}[direct predecessor \cite{a:rtssgradient}] \label{def:neighbour}
  An event $e'=(\ID',t')$ is a \emph{direct predecessor} (or \emph{neighbour} for short) of an event $e=(\ID,t)$, denoted by $e' \neigh e$, if the message broadcast by $e'$ was the last from $\ID'$ able to reach $\ID$ before $e$ occurred (and was not discarded by $i$ as an obsolete message).
\end{definition}

It follows that if $e'$ is a neighbour of $e$, then $e'$ has to happen right before $e$, but not too long time ago (otherwise the message would have been discarded) or too far away (otherwise the message would not be received): thus, the neighbouring relation typically reflects spatial proximity. However, it could also be a logical relationship (e.g., connecting master devices to slave devices independently of their position), in which case the ``far away'' requirement would be measured through the logical network topology.

Furthermore, notice that the relation $\neigh$ on events forms a direct acyclic graph (DAG) among events, since cycles would correspond to a closed timelike curve. Hence, the $\neigh$ relation is time-driven and anti-symmetric, unlike spatial-only neighbouring (which is usually symmetrical).

\begin{figure}[t]
	\centering
	\includegraphics[scale=1.5]{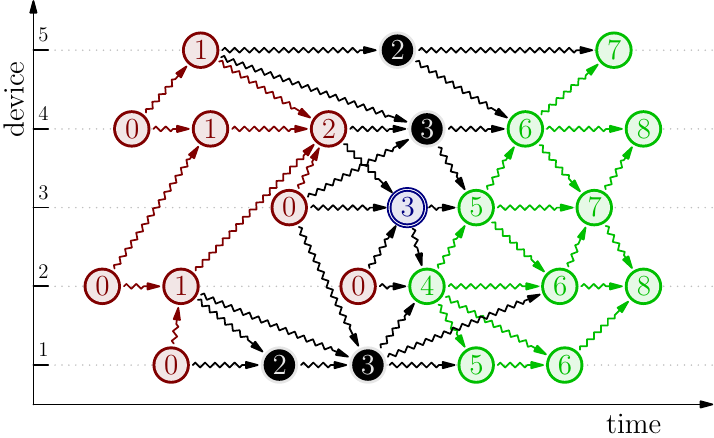}
	\caption{Representation of a field evolution of integers together with its underlying event structure (neighbouring). Past events of the circled blue event $3$ are depicted in red, future events in green, concurrent events in black. This field evolution models the computation in each event of the longest preceding chain of events, obtainable locally by taking the maximum of the neighbour counters increased by $1$.} \label{fig:structure}
\end{figure}

\begin{definition}[causality \cite{a:rtssgradient}]
	The \emph{causality} partial order $e' \leq e$ on events is the transitive closure\footnote{Thus, $e' \leq e$ iff there exists a (possibly empty) sequence $e_1, \ldots, e_n$ of events such that $e' \neigh e_1 \neigh \ldots \neigh e_n \neigh e$.} of $\neigh$.
\end{definition}

The causality relation defines which events constitute the past, future or are concurrent to any given event. A set of events together with a neighbouring relation determines an \emph{event structure}, represented in Figure \ref{fig:structure}. Notice that we do not assume that a global clock is available, nor that the scheduling of events follow a particularly regular pattern. This choice is dictated by the need to apply the field calculus to the broadest possible class of problems, as further restrictions can always be added without hassle.

Using the formalism of event structures, we can abstract the data manipulated by a field calculus program as a whole distributed space-time \emph{field  evolution} $\Phi$, mapping individual events $e$ in an event structure $E$ to data values $\anyvalue$ (see Figure \ref{fig:structure}). Similarly, we can understand an ``aggregate computing machine'' as a device manipulating these field evolutions, and abstract is as a function mapping input field evolutions to output field evolutions.

\subsection{The programming language}

The syntax of field calculus is presented in Figure~\ref{fig:syntax}---the overbar notation
$\overline\e$ is a shorthand for sequences of elements, and multiple overbars are intended to be
expanded together, e.g., $\overline\e$ stands for $\e_1, \ldots, \e_n$ and 
$\envmap{\overline\ID}{\overline\lvalue}$ for
$\ID_1 \mapsto \lvalue_1, \ldots, \ID_n \mapsto \lvalue_n$.  
The keywords $\nbrK$ and $\repK$ correspond to the two peculiar constructs of
field calculus, responsible of interaction and field dynamics, respectively; while
$\defK$ and $ \ifK$ correspond to the standard function definition
and the branching expression constructs.

\begin{figure}[!t]
\centering
\centerline{\framebox[\linewidth]{$
        \begin{array}{lcl@{\hspace{8mm}}r}
                \PROGRAM & \BNFcce & \overline{\FUNCTION}  \; \e
                &{ \mbox{\footnotesize program}}
                \\[3pt]
                \FUNCTION & \BNFcce &  \defK \,\; \fname (\overline{\xname}) \; \{ \e \}
                &{ \mbox{\footnotesize function declaration}}
                \\[3pt]
                \e & \BNFcce &  \xname \;\BNFmid\; \funvalue(\overline\e) \;\BNFmid\; \anyvalue \;\BNFmid\; \ifK (\e) \{\e\} \{\e\} \;\BNFmid\; \nbrK\{\e\} \;\BNFmid\; \repK(\e)\{ (\xname) \toSymK \e \}
                &{ \mbox{\footnotesize expression}}
                \\[3pt]
                \funvalue & \BNFcce &  \fname \; \BNFmid \; \bname
                &{ \mbox{\footnotesize function name}}
                \\[3pt]
                \anyvalue & \BNFcce &  \lvalue \; \BNFmid \; \fvalue
                &{ \mbox{\footnotesize value}}
                \\[3pt]
              \lvalue & \BNFcce &  \dcOf{\dc}{\overline\lvalue} 
                &{ \mbox{\footnotesize local value}}
                \\[3pt]
                \fvalue & \BNFcce &  \envmap{\overline\ID}{\overline\lvalue}
                &{ \mbox{\footnotesize neighbouring field value}}
                \\[3pt]
        \end{array}
        $}
}
\caption{Syntax of the field calculus language.}
\label{fig:syntax}
\end{figure}

A program  $\PROGRAM$ is the declaration of a set of functions $\overline\FUNCTION$ of the kind
``$ \defK \,\; \fname (\xname_1,\ldots,\xname_n) \; \{ \e \}$'', and a main expression $\e$ that is the one executed at each computation round, as well as the one considered (in the global viewpoint) as the overall field computation.
An expressions $\e$ can be: 
\begin{itemize}
        \item A \emph{variable} $\xname$, used e.g. as formal parameter of functions.
        \item A \emph{value} $\anyvalue$, which can be of the following two kinds: 
        \begin{itemize}
       \item
            A \emph{local value} $\lvalue$, with structure $\dcOf{\dc}{\overline\lvalue}$ or simply $\dc$ when $\overline\lvalue$ is empty (defined via data constructor $\dc$ and arguments $\overline\lvalue$), can be, e.g., a Boolean (\lstinline|True| or \lstinline|False|), 
              a number, a string, or a structured value (e.g., a pair \lstinline|Pair(True,5)|).
            \item  
                A \emph{neighbouring (field) value} $\fvalue$ 
                that  associates neighbour devices $\ID$ to local values $\lvalue$, e.g., it could be the neighbouring value of distances of neighbours---note that neighbouring field values are not part of the surface syntax, they are produced at runtime by evaluating 
expressions, as described below.
        \end{itemize}
              \item A function call $\funvalue(\overline\e)$, where $\funvalue$ can be of two kinds: a \emph{user-declared function} $\fname$
    (declared by the keyword $\defK$, as illustrated above)
           or a \emph{built-in function} $\bname$, such as a mathematical or logical operator, a data structure operation, or a function returning the value of a sensor.
         \item A branching expression $\ifK (\e_1) \{\e_2\} \{\e_3\}$, used to
           split field computation in two isolated sub-networks, where/when $\e_1$ evaluates to  \lstinline|True| or \lstinline|False|: the result is computation of $\e_2$ in the former area, and $\e_3$ in the latter.
           
         \item An $\nbrK$-expression $\nbrK\{\e\}$, use to create a neighbouring field value mapping neighbours to their latest available result of evaluating $\e$. In particular, each device $\ID$:
           \begin{enumerate}
           \item broadcasts (together with its state information) its value of $\e$ to its neighbours,
           \item evaluates the expression into a neighbouring field
             value $\fvalue$ associating to each neighbour $\ID'$ of
             $\ID$ the latest evaluation of $\e$ at $\ID'$.
           \end{enumerate}
           Note that the the evaluation by a device $\ID$
           of an $\nbrK$-expression within a branch of some
           $\ifK(\e_1)\dots$ expressions, is affected only by the
           neighbours of $\ID$ that, during their last computation
           cycle, evaluated the same value for $\e_1$.
         \item A $\repK$-expression $\repK(\e_1)\{(\xname) \toSymK{} \e_2\}$
           models evolution through time, by returning the value
           of the expression $\e_2$ where each occurrence of $\xname$
           is replaced by the value of the $\repK$-expression at the
           previous computation cycle---or by $\e_1$ if the
           $\repK$-expression has not been evaluated in the previous
           computation cycle.
\end{itemize}
The meaning of a field calculus program can be defined through a denotational and an operational semantics, both thoroughly studied in \cite{Viroli:HFC-TOCL}. The denotational semantics $\denotexp{\e}{}{E}$ maps an expression $\e$ to a field evolution $\Phi = \denotexp{\e}{}{E}$ on a given event structure $E$ (see Section \ref{ssec:model}), and is compositional meaning that $\denotexp{\funvalue(\e)}{}{E} = \denotexp{\funvalue}{}{E}(\denotexp{\e}{}{E})$.

Alternatively, an operational semantics of rounds in a network can be given in terms of a transition system $\nettran{N}{\act}{N'}$ between network configurations $N$, where $\act$ is either $\envact$ to model any environment change, or a device identifier to represent a device computation. These computations are in turn modelled by a local judgement $\bsopsem{\ID}{\Theta}{\sigma}{\e}{\theta}$ to be read as ``expression $\e$ evaluates to $\theta$ on device $\ID$ with respect to sensor values $\sigma$ and neighbours' data $\Theta$'', where $\theta$ is the structure of values obtained from the evaluation of every sub-expression of $\e$, and $\Theta$ is a map $\envmap{\overline\ID}{\overline\theta}$ from neighbour device identifiers to the last $\theta_\ID$ which was received from them by the current device.

\begin{example}[hop-count distance] \label{ex:hopcount}
In order to give an intuition of the behaviour of a field calculus program, consider the following function, where \lstinline|minHood| selects the minimum element in the range of a numeric field $\fvalue$, and \lstinline|mux| is a classic ``multiplexer'' operator selecting its second or third argument depending on the truth value of the first (overloaded to apply pointwise on fields).
\begin{lstlisting}
def hopcount(source) {
  rep (infinity) { (c) => mux(source, 0, minHood(nbr{c+1})) }
}
\end{lstlisting}
The \lstinline|hopcount| functions computes the number of hops required to reach a node where \lstinline|source| is true: it is zero in sources, and equal to the minimum count of a neighbour incremented by one in non-source nodes. 
\end{example}

\begin{remark}[sample code]
	In practical implementations of the field calculus, the language is often extended to include additional features improving code readability. In Section \ref{sec:FCMonitor} we shall use some of them, in particular:
	\begin{itemize}
		\item The traditional \lstinline|let x = e_1 in e_2| construct, which can be thought as a shorthand for the expression \lstinline|f(e_1, y_1,...,y_n)| given the definition \lstinline|def f(x, y_1,...,y_n) {e_2}|, where \lstinline|y_1|, \ldots, \lstinline|y_n| are the variables occurring free in \lstinline|e_2|.
		\item The notation \lstinline|[e_1,...,e_n]|, representing tuple creation \lstinline|Tuple(e_1,...,e_n)|.
		\item The multi-valued $\repK$ construct \lstinline|rep (v_1,...,v_n) {(x_1,...,x_n) => e_1,...,e_n}|, as a shorthand for the following.
\begin{lstlisting}
rep ([v_1,...,v_n]) { (t) =>
  let x_1 = 1st(t) in ... let x_n = nth(t) in [e_1,...,e_n]
}
\end{lstlisting}
	\end{itemize}
\end{remark}

\section{Implementing monitors in field calculus}\label{sec:GeneralMonitor}

Inspired by Francalanza et al.~\cite{Francalanza2018}, we frame our discussion by considering a distributed monitoring setting where:
\begin{enumerate}
\item
The system under analysis comprises a number of subsystems, identified by \emph{processes} $\Pi$,  that execute independently and  
might interact (i.e., synchronize or communicate) via the underlying communication platform.
\item
The set of processes is partitioned across locations $\lambda$, i.e., each process $\Pi$ is located at exactly one location, denoted by $\locationOf(\Pi)$.
Two processes $\Pi$ and $\Pi'$ are \emph{local} to one other if and only if  $\locationOf(\Pi)=\locationOf(\Pi')$, and \emph{remote} otherwise.
Processes may interact with both local and remote processes (usually remote communication is more expensive than local communication). Notable cases are when one of the following two conditions holds:
\begin{enumerate}
\item
There is just one location (i.e., all the processes are local); 
\item
At each location there is exactly one process (i.e., all processes are remote).
\end{enumerate}
\item
Each location hosts a number of \emph{local traces} $\tau$,
 each trace consists of a total ordered set of \emph{events}, and each event describes a discrete computational step of a process at the location that hosts the trace.
A trace may contain events of different process. Notable case are when one or two of the following conditions holds:
\begin{enumerate}
\item
each trace contains events of a single process (i.e., each trace belongs to a single process), or 
\item
for each process there is exactly one trace (containing the events of the process).
\end{enumerate}
\item
Monitoring is performed  by computation entities, identified by \emph{monitors} $M$,  that check properties of the system under analysis by 
analysing the traces.  Similar to processes each monitor is hosted at a given location and may communicate with other (local or remote) monitors. 
Notable cases are  when there is exactly one monitor for each:
\begin{enumerate}
\item
location,
\item
process, or
\item
trace.
\end{enumerate}
\end{enumerate}

In runtime verification monitors are generated from formal specifications.
In the following we illustrate, by means of simple examples,  how the field calculus can be used to implement distributed monitors.
Our aim is to pave the way towards generating field calculus distributed monitors from suitable formal specifications.
%
%
We consider the following setting:
\begin{itemize}
\item
Each monitor is implemented by a field calculus program running on a dedicated (virtual or physical) device.
\item
Each local trace is mapped to a sensor.
\item
The $\nbrK$ construct comes in two forms: 
\begin{itemize}
\item
\texttt{nbrLocal}, for communication with local devices (i.e., if $\ID_1$ is a neighbour of $\ID_2$ then they are at the same location), and
\item
\texttt{nbrRemote}, for communication with remote devices (i.e., if $\ID_1$ is a neighbour of $\ID_2$ then they are at different locations).
\end{itemize}
\item
Each device  $\ID$ is awaken whenever:
\begin{itemize}
\item
a new event arrives on one of the sensors of the devices, or
\item
a new \emph{different} message arrives from a (local or remote) neighbour $\ID'$;\footnote{Note that if the new message
 is equal to the last message received from  $\ID'$ then the device  $\ID$ is not awaken.}
\end{itemize}
provided that a minimum time span $T$ has elapsed from the previous evaluation cycle.
\end{itemize}
Moreover, for simplicity, we also assume that both conditions 3.a and 3.b hold. 
%
%
In the next two subsections we  present examples in the context of the  ``local monitor only'' and of the ``remote monitors only''  assumptions, respectively.

\subsection{Local monitors only} \label{sec:LocalMonitorsOnly}

In this subsection we assume that condition 2.a (given at the beginning of Section~\ref{sec:GeneralMonitor}) holds, that is, every process is local.
We consider two smart home scenarios, in which processes are assumed to be local through either: \textit{(i)} physically wired connections; or \textit{(ii)} short-range efficient wireless communication, as the one expected by upcoming 5G standards. In this setting, the network topology can be full, that is, every node communicates with every other node.

In the first scenario, we want to monitor the following property: \emph{air conditioning and lights are on when the room is not empty}. In order to express this property, we assume that the following 0-ary built-in operators (with corresponding traces) are given:
\begin{itemize}
	\item \lstinline|lights|: an optional Boolean value, which is true if the lights are on, false if they are off and null in nodes not controlling the lights.
	\item \lstinline|people|: an optional Boolean value, depending on whether the node is sensing the presence of nearby people (if sensing is available).
\end{itemize}
This first property can then be expressed through the following program:
\begin{lstlisting}
lights() == null || lights() == anyHood(nbrLocal{people() == true})
\end{lstlisting}
where \lstinline|anyHood| is a built-in function that given a Boolean field $\fvalue$, returns true if and only if at least one element in the range of $\fvalue$ is true. The monitoring property holds in nodes not controlling lights (i.e., when \lstinline|lights| is \lstinline|null|), or when the lights are on if and only if \lstinline|people| is true in some communicating node, capturing the required idea.

In the second scenario, we want to monitor the following property: \emph{if the volume of the stereo is above a certain threshold, every node should rapidly agree on alerting the stereo to lower its volume}. In order to express this property, we assume that the following 0-ary built-in operators (with corresponding traces) are given:
\begin{itemize}
	\item \lstinline|level|: the volume level of the stereo, or \lstinline|0| in nodes not controlling the stereo.
	\item \lstinline|alert|: an optional Boolean value, depending on whether the node is sensing excessive noise, which is null if no sensing is available.
\end{itemize}
This second property can then be expressed through the following program:
\begin{lstlisting}
def roundsince(condition) {
  rep (0) { (x) => if (condition) {0} {x+1} }
}
roundsince(allHood(nbrLocal{alert() != false}) || level() <= THRESHOLD) < DELAY
\end{lstlisting}
where \lstinline|THRESHOLD|, \lstinline|DELAY| are given constants and \lstinline|allHood| is a built-in function that given a Boolean field $\fvalue$, returns true if and only if every element in the range of $\fvalue$ is true. Function \lstinline|roundsince| counts the number of rounds elapsed since the last time \lstinline|condition| was true. The monitoring property holds provided that no more than \lstinline|DELAY| turns elapsed since when the volume was below \lstinline|THRESHOLD| or all nodes agreed on alerting.

\subsection{Remote monitors only} \label{sec:RemotelMonitorsOnly}

In this subsection we assume that condition 2.b (given at the beginning of Section~\ref{sec:GeneralMonitor}) holds, that is, every process is remote. In this case, it is no longer realistic to assume a full communication topology; instead, we shall have few neighbours for every node to reduce the number of needed communications. This may not make a difference in case the property to monitor is fully local, as by the first example discussed in Section \ref{ssec:examples} which may be written through the following specification, where \lstinline|nbrVector| is a returns the field of vectors to neighbours, \lstinline|direction| is the quantity to be monitored and \lstinline|angle| computes the relative angle between two vectors.
\begin{lstlisting}
allHood(-60 < angle(nbrVector(), direction()) < 60 &&
        -60 < angle(-nbrVector(), nbr{direction()}) < 60)
\end{lstlisting}
However, when properties to monitor are not fully local, we may require a ``data collection'' routine to ensure effective spatial quantification (e.g., checking whether a property is true for all devices). This can be accomplished in field calculus through the \emph{collection} building block, which is here instantiated for spatial quantification with the help of the result \lstinline|count| of the simple distance estimation routine \lstinline|hopcount| described in Example \ref{ex:hopcount}.
\begin{lstlisting}
def everywhere(property, count) {
  rep (false) { (p) =>
    allHood(mux(nbrRemote{count} > count, nbrRemote{p}, property))
} }
def somewhere(property, count) {
  rep (false) { (p) =>
    anyHood(mux(nbrRemote{count} > count, nbrRemote{p}, property))
} }
\end{lstlisting}
The \lstinline|everywhere| and \lstinline|somewhere| functions check the validity of a property in nodes with a higher count, so that their value in the source should correspond to the intended result. More efficient collection \cite{a:collection,viroli:selfstabilisation} and distance computation algorithms \cite{a:ultgradient,a:scpgradient} may be used in practical systems to implement those same functions: in this paper, we opted for the simplest implementations instead for sake of readability.

With the help of those functions, we can translate both scenarios in Section \ref{sec:LocalMonitorsOnly} to a remote-only setting. For the first scenario, we may want to check that an electronic system is on when some people are present in a large building, which can be accomplished by the following code.
\begin{lstlisting}
lights() == null || lights() == somewhere(people() == true, hopcount(lights()!= null))
\end{lstlisting}
For the second scenario, we may want to check that every area of such a building is alerted for evacuation after some dangerous event has been detected, which can be accomplished by the following code.
\begin{lstlisting}
roundsince(everywhere(alert() != false, hopcount(level() != 0)) || 
           level() <= THRESHOLD) < DELAY
\end{lstlisting}
In both scenarios, we compute hop-count distances from controller nodes (which are reasonably unique), and use these distances to guide aggregation.

\section{Monitoring field calculus programs}\label{sec:FCMonitor}

In case the distributed program to be monitored is a field calculus program, further opportunities arise from the ability of instrumenting the monitor code within the original algorithm, and possibly implementing feedback loops between them. Inspired by the second motivating example presented in Section \ref{ssec:examples}, we consider the following \lstinline|channel| routine building on the \lstinline|hopcount| function presented in Example \ref{ex:hopcount}.

\begin{lstlisting}
def broadcast(value, count) {
  rep (value) { (oldval) =>
    mux( count == 0, value, 2nd(minHood(nbr{[count, oldval]})) )
} }
\end{lstlisting}
\begin{lstlisting}
def elliptic-channel(sourcecount, destcount, width) {
  let sourcedest = broadcast(sourcecount, destcount) in
  sourcecount + destcount <= sourcedest + width
}
def channel(value, source, dest, width) {
  let sourcecount = hopcount(source) in
  let destcount   = hopcount(dest) in
  let inarea = elliptic-channel(sourcecount, destcount, width) in
  if (inarea) { broadcast(value, sourcecount) } { value }
}
\end{lstlisting}

The \lstinline|broadcast| function spreads a \lstinline|value| from a source generating a certain hop-count distance (\lstinline|count|) outwards: every device selects the provided value only if it is the source (\lstinline|count == 0|), otherwise it selects the value of the neighbour with minimal \lstinline|count|. Function \lstinline|elliptic-channel| defines a roughly elliptic area with foci in a source and destination and given \lstinline|width|, by comparing the sum of distances from the current location to the source and destination with the distance between the source and destination themselves (obtained by broadcasting from the destination the value of the distance to the source). Finally, function \lstinline|channel| uses the above functions to broadcast a value in the area selected by \lstinline|elliptic-channel|.

In order for the communication to be reliably performed, the \lstinline|width| parameter has to be carefully tuned, depending also on the network characteristics. Thus, it is crucial to monitor the effectiveness of the choice, as performed by the following functions, where \lstinline|sumHood| computes the sum of a numeric field $\fvalue$, \lstinline|min| computes the minimum between two numbers, and \lstinline|myID| returns the identifier of the current device.

\begin{lstlisting}
def samevalue(value, count) {
  let num,id = rep (1,myID()) { (num,id) =>
    sumHood(mux(nbr{id} == myID(),  num, 0))+1,
    2nd(minHood( mux(nbr{value} == value, nbr{[count,myID()]}, [infinity, myID()]) ))
  }) in
  broadcast(num, if (id == myID()) {0} {count})
}
def monitor(sourcecount, destcount, minw, maxw) {
  let w = min(samevalue(sourcecount,destcount), samevalue(destcount,sourcecount)) in
  if (w > maxw) {HIGH} {if (w < minw) {LOW} {OK}}
}
\end{lstlisting}

Function \lstinline|samevalue| computes the number of devices holding the same value for \lstinline|value| in devices with the lowest possible \lstinline|count|: every device collects partial estimates \lstinline|num| from neighbours who selected it in \lstinline|id|, and selects in \lstinline|id| the neighbour with the same \lstinline|value| and lowest possible \lstinline|count|. The \lstinline|num| computed by the device with the lowest possible \lstinline|count| is then broadcast to others (since devices with lowest possible \lstinline|count| select themselves as \lstinline|id|).
The \lstinline|monitor| then uses function \lstinline|samevalue| to estimate the cross-section from both points of view of the source and destination, considering the minimum among them: a status is finally returned depending on whether this estimates fall above, below or within the required interval.

This monitor, if run within the area selected by \lstinline|elliptic-channel|, can estimate whether the channel is properly established. Furthermore, it can be instrumented within the channel function to obtain an auto-adjusting channel as in the following.

\pagebreak
\begin{lstlisting}
def adjusting-channel(value, source, dest, minw, maxw) {
  let sourcecount = hopcount(source) in
  let destcount   = hopcount(dest) in
  let inarea = 1st(rep (False, maxw) { (oarea, owidth) =>
    let narea = elliptic-channel(sourcecount, destcount, owidth) in
    let status = if (narea) {monitor(narea, minw, maxw)} {OK} in
    narea, if (status == OK) {width} {if (status == LOW) {owidth+1} {owidth-1}}
  }) in
  if (inarea) { broadcast(value, sourcecount) } { value }
}
\end{lstlisting}

This function increases or decreases the width by 1 according to the status returned by the monitor. Furthermore, it does so independently in \emph{every device} of the network, allowing the shape of the channel to adjust to the network local peculiarities (instead of the fixed elliptical shape of the traditional \lstinline|channel|).

\section {Conclusion}\label{sec:Conclusion}

In this position paper we have illustrated, by means of simple examples, how the field calculus can be used to implement distributed monitors in different settings.  In particular, we have provided examples of local and remote monitors, and an example of a field calculus program within which the monitor can be instrumented providing the algorithm with an additional auto-correcting power.

In future work we would like to investigate how field calculus expressions, e.g.\ using the $\nbrK$-construct, could be used in conjunction with a specification language like LTL; and possibly be automatically generated by a logical language. This would allow us to write properties along the lines of ``Eventually, all my neighbours\ldots'' or ``Some neighbour will always \ldots''.

\subsection*{Acknowledgements}
               
This work has been partially supported by the European Union's Horizon 2020 research and innovation programme under project COEMS (\url{www.coems.eu}, 
grant agreement no.~732016),  project HyVar (\url{www.hyvar-project.eu}, grant agreement no.~644298) and ICT COST 
Action IC1402 ARVI (\url{www.cost-arvi.eu}). 
We thank the anonymous VORTEX 2018 reviewers for insightful comments and suggestions for improving the presentation.

\nocite{*}
\bibliographystyle{eptcs}
\bibliography{biblio}

\end{document}